\documentclass{iau} 
\usepackage{graphicx}
\usepackage{natbib}
\usepackage{hyperref}

\title[Hunting for red supergiant binaries] 
{Hunting for red supergiant binaries:\\
UVIT photometry of the SMC}

\author[Lee R. Patrick \textit{et al.}]   
{Lee R. Patrick$^{1,2, 3}$,
David~Thilker,$^{4}$,
Danny~Lennon,$^{5, 6}$,
Luciana~Bianchi,$^{4}$,
Abel~Schootemeijer,$^{7, 8}$,
Ricardo~Dorda,$^{5, 6}$,
Norbert~Langer$^{7, 8}$,
\and Ignacio Negueruela$^{2}$}

\affiliation{$^1$Departamento de Astrof\'{\i}sica, Centro de Astrobiolog\'{\i}a,
(CSIC-INTA), Ctra. Torrej\'on a Ajalvir, km 4,  28850 Torrej\'on de Ardoz,
Madrid, Spain\
\\[\affilskip]
$^2$Departamento de F\'{\i}sica Aplicada, Universidad de Alicante, E-03690 San Vicente del Raspeig, Alicante, Spain
\\[\affilskip]
$^3$School of Physical Sciences, The Open University, Walton Hall, Milton Keynes MK7 6AA, UK
\\[\affilskip]
$^4$Department of Physics and Astronomy, Johns Hopkins University, 3400 N. Charles Street, Baltimore, MD 21218, USA
\\[\affilskip]
$^5$Instituto de Astrof\'isica de Canarias, E-38205 La Laguna, Tenerife, Spain
\\[\affilskip]
$^6$Universidad de La Laguna, Dpto. Astrof\'isica, E-38206 La Laguna, Tenerife, Spain
\\[\affilskip]
$^7$Argelander-Institut f\"ur Astronomie, Universit\"at Bonn, Auf dem H\"ugel 71, 53121, Bonn, Germany
\\[\affilskip]
$^8$Max-Planck-Institut für Radioastronomie, Auf dem H\"ugel 69, 53121, Bonn, Germany
}

\pubyear{2022}
\volume{361}  
\jname{Massive Stars Near and Far}
\editors{N.~St-Louis, J.~S.~Vink \& J.~Mackey, eds.}
\begin{document}

\maketitle

\begin{abstract}
We present UVIT/Astrosat UV photometry of the RSG population of the Small Cloud galaxy (SMC).
As RSGs are extremely faint in the far-UV, these observations directly probe potential companion stars.
From a sample of 861 SMC RSGs, we find 88 have detections at far-UV wavelengths: a clear signature of binarity.
Stellar parameters are determined for both components, which allows us to study - for the first time - the mass-ratio ($q$) distribution of RSG binary systems.
We find a flat mass-ratio distribution best describes the observations up to M$_{\rm RSG}\sim$15M$_\odot$.
We account for our main observing bias (i.e. the limiting magnitude of the UVIT survey) to determine the intrinsic RSG binary fraction of 18.8\,$\pm$\,1.5\,\%, for mass-ratios in the range $0.3 < q < 1.0$ and orbital periods approximately in the range $3 < \log P[{\rm days}] < 8$.
\end{abstract}

\firstsection 
\section{Introduction}

Multiplicity is a key uncertainty in the evolution and eventual fate of massive stars~\citep{Langer12}. 
In recent times there is a growing realisation that the majority of massive stars are born with a close companion~\citep[e.g.][]{2012Sci...337..444S}.
Close companions may interact and merge~\citep{2017ApJS..230...15M,2020A&A...638A..39L,2022A&A...659A..98S}, which results in blue straggler stars~\citep{2014ApJ...780..117S} and exotic stellar systems (in originally higher-order multiple systems).
The result of such evolution on the red supergiant star population, which is either one of, or the final, evolutionary stage for the majority of massive stars (between $\sim$8 and 40\,M$_\odot$), is not well constrained observationally. 

Theoretically, when a star within a close binary system evolves off the main sequence towards the RSG phase, an interaction occurs, which either results in a merger event or some form of envelope stripping. The majority of such interactions are expected to result in an evolutionary path that avoids a RSG phase~\citep{2008MNRAS.384.1109E}.

Stars that have been regenerated in a merger event return to the main-sequence~\citep{2019Natur.574..211S} and subsequently evolve towards the RSG phase where they are expected to appear as red analogues to blue-straggler stars, so called \textquoteleft red stragglers\textquoteright~\citep{b19},
where they can account for up to 50\,\% of RSG populations in clusters~\citep{2019MNRAS.486..266B,b19,2020A&A...635A..29P}.

Red supergiant stars are very red and luminous, therefore, massive binary systems where one component is a RSG are likely to be dominated by the appearance of the RSG at optical and infrared wavelengths.
Such systems are also necessarily longer orbital period systems, where the minimum allowed orbital periods for a RSG are between 100 and 1000 days.
As the lifetime of the RSG phase is much shorter than that of the main-sequence, one expects this to place a lower limit on the mass of a potential companion~\citep{2018AJ....155..207N}, although this lower limit is not well constrained observationally as lower mass, lower mass-ratio systems are typically most difficult to detect (see Section~4 for more discussion). 

The impact of binary systems on stellar populations is important to study binary physics and to predict the outcomes of binary evolution.
For example, the increased the line-of-sight velocity distribution of young clusters is seen as an indicator of the level of binarity~\citep{2010MNRAS.402.1750G}, however, this is not observed in older clusters dominated by RSGs~\citep{2016MNRAS.458.3968P,2020A&A...635A..29P}, which is an indication that the multiplicity fraction of RSGs is lower than in earlier evolutionary phases.

To identify RSG binary systems there are typically two approaches: 

\begin{enumerate}
    \item Detect radial velocity variations that cannot be explained as stellar activity from a single star.
    Given the long orbital periods, long-baseline observations are a requirement for the success of this technique.
    \citet{burki} first performed such a study in the Galaxy, detecting 19 K- and M-type RSG binary systems. 
    \citet{p19} and~\cite{2020A&A...635A..29P} applied this technique to star clusters in both the Large and Small Cloud galaxies (MCs), however, these authors could not detect individual systems.
    \citet{2021MNRAS.502.4890D} calibrated a large baseline set of inhomogeneous observations in the MCs to the Gaia DR2 radial velocity rest frame.
    These authors were able to identify individual systems, but couldn't correct for the various observational biases to determine intrinsic multiplicity fractions. 
    
    \item Detect a signature of the companion in photometric or spectroscopic observations.
    \citet{2018AJ....155..207N} developed a method that exploits widely-available optical photometry to identify a blue excess that is attributed to the presence of a companion.
    This method has been refined in a series of studies that identifies RSG binary systems in several Local Group galaxies~\citep{2020ApJ...900..118N,2021ApJ...908...87N}.
    From their observations, these authors determine the intrinsic binary fraction of RSGs using detailed simulations.
\end{enumerate}

Table~\ref{tb:RSG-bf} provides an overview of the results of these various studies.

In this article we summarise the results of a search for RSG binary systems with newly available UV photometry from the UVIT/Astrosat SMC survey (Thilker et al. in prep.).
UV photometry has the unique benefit that an early-type companion contributes close to 100\% of the flux at such wavelengths and, as such, the companions of a large sample of RSG binary systems can be studied directly, which is the first time that this has been possible.

\section{Methods}

As our aim is to identify RSG binary systems we first construct a catalogue of 861 RSGs in the SMC, which is based on the source catalogue of \citet{2020A&A...639A.116Y}. 
This is a list of high-probability SMC RSGs, based on the photometric classification at multiple wavelengths and astrometric data provided by Gaia EDR3~\citep{2021A&A...649A...1G}.

Thilker et al. (in prep.) presented the UVIT survey of the SMC in the F172M filter, which has a central wavelength of 1717\,\AA.
This SMC-wide survey\footnote{of the SMC $\ldots$} is currently 75\,\% complete.
This survey has a limiting magnitude of $m_{F172M}=20.3$\,ABmag with an astrometric precision of around 0.1" and a 4x better spatial resolution than GALEX.

We identify RSG binary systems by cross-matching the two catalogues. 
To determine the contribution from chance alignments, we follow the method of~\citet{2020ApJS..250...36B} and use a control catalogue, which is the original RSG catalogue offset by 10" in declination.
The results of this analysis show that outside of a cross-match distance (XMD) of 0.5", the probability of chance alignments dominates the matches, which highlights the need for precision astrometry in such studies.
To minmise the possibility of chance alignments we select a maximum XMD of 0.4".
This results in 88 matches, which we interpret as genuine RSG binary systems.


\section{Stellar parameters of both components}
\begin{figure}
\begin{center}
 \includegraphics[width=0.65\textwidth]{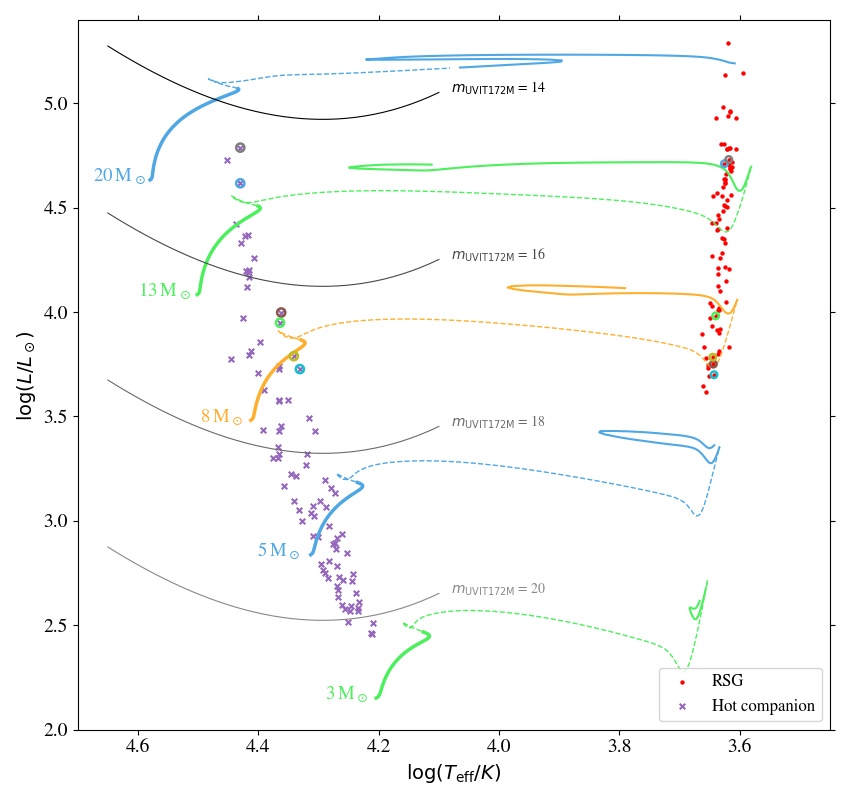} 
 \caption{Hertzsprung--Russel diagram showing the RSGs and their companions. The coloured lines show stellar evolutionary tracks that are based on those in~\citet{2019A&A...625A.132S}.
 The solid grey lines highlight lines of constant UVIT magnitudes in the stellar models. Rings of different colours highlight the six systems that have a mass-ratio greater than 1.}
   \label{fig:HRD}
\end{center}
\end{figure}

We utilise photometry to determine the stellar parameters of both components within each of the 88 RSG binary systems and for the entire RSG catalogue. 
In the following analysis we assume a single extinction law for all targets, which is characterised with $A_V=0.35$ and the SMC bar reddening law of~\citet{2003ApJ...594..279G}.
See~\citet{2022MNRAS.513.5847P} for further details.
Figure~\ref{fig:HRD} displays the stellar parameters determined for the stars within binary systems.

The luminosities of the RSG component are determined from $K$-band photometry using the SMC calibrations of~\citet{2013ApJ...767....3D}.
The RSG effective temperatures are determined using a calibration to the de-reddened $J-K$ colours~\citep{2021MNRAS.502.4890D}.
This calibration is based on the RSG effective temperatures determined for SMC RSGs in~\citet{2018MNRAS.476.3106T}.
A discussion of SMC RSG effective temperatures is provided in Appendix~B of~\citet{2022MNRAS.513.5847P}. 
We determine the evolutionary masses and ages of the RSG via a comparison with MESA models computed for the SMC by~\citet{2019A&A...625A.132S}.
These models use a mass-dependent overshooting parameter and a semiconvection parameter of 10.
We determine the mass and age of the RSG by comparing the observed luminosity of the RSG with the luminosity of the mid-point of the helium burning lifetime of the models.

To determine the stellar parameters of the companions we make the simplifying assumption that each companion is a single, main-sequence star that is the same age as the RSG.
We determine the luminosity of the companion using the UVIT photometry, assuming a bolometric correction taken from the MIST models~\citep{2016ApJ...823..102C,2016ApJS..222....8D}.
At these wavelengths there is effectively no contribution from the RSG. 
To determine the mass and effective temperature of the companion, we compare the luminosity of the companion with the same MESA models as for the RSG.



\section{Key results and discussion}
\begin{figure}
\begin{center}
 \includegraphics[width=0.48\textwidth]{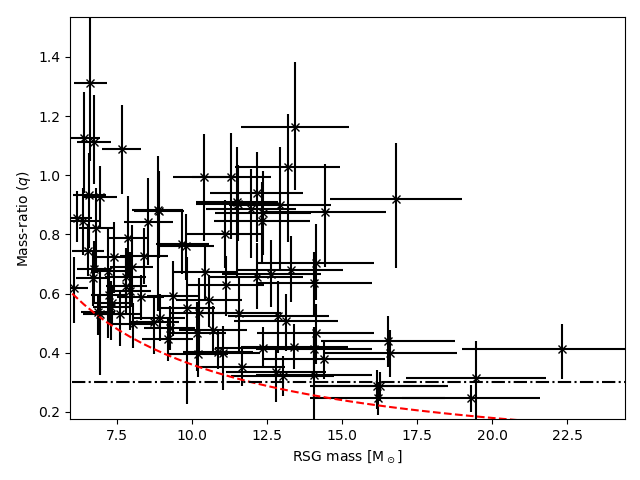} 
 \includegraphics[width=0.50\textwidth]{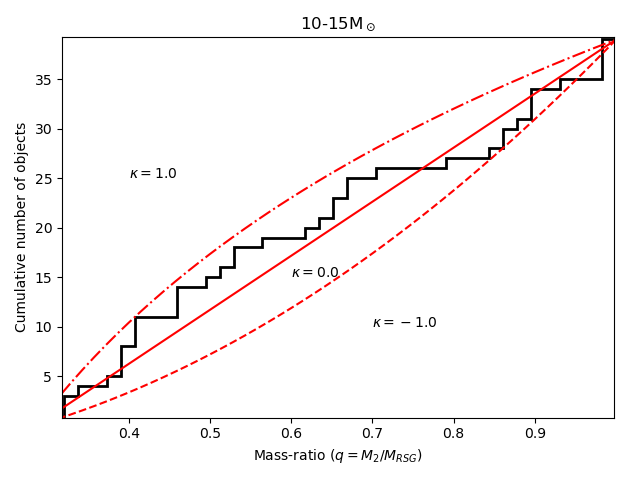} 
 \caption{Left: Mass-ratio distribution as a function of RSG mass. The red dashed line shows the mass-ratio dependent observing bias. The black dot-dashed line shows a mass-ratio limit of $q = 0.3$.
 Right: Mass-ratio cumulative distribution function for RSGs masses between 10 and 15\,M$_\odot$.}
  \label{fig:q-figs}
\end{center}
\end{figure}
We determine mass-ratios for all of the 88 RSG binary systems detected.
Figure~\ref{fig:q-figs} shows the mass-ratio distribution as a function of RSG mass and the cumulative distribution function for RSG binary systems for RSGs within the mass range 10 $<$M/M$_\odot<$15. 
The left panel highlights the principle observing bias of this survey, which is that the limiting magnitude of the UVIT survey results in a mass-ratio dependent observing bias such that low mass-ratio systems go undetected at low RSG masses. 

To minimise the contribution of the observing bias on the mass-ratio distribution, we concentrate on RSG masses in the range 10 $<$M/M$_\odot<$15 and mass-ratios in the range $0.3 < q < 1.0$.
RSG binary systems in this parameter range are best described with a flat mass-ratio distribution.
We also find a lack of high-mass ratio systems at above 15\,M$_\odot$, which, in principle, should be the easiest systems to detect given our observations. 

We account for observational biases by simulating the observed population of RSG binaries assuming a flat mass-ratio distribution with the aim of effectively filling in the area of the left panel of Figure~\ref{fig:q-figs}, that could not be observed.
From 100\,000 simulations we determine that we have missed on average 17.7\,$\pm$\,4.4 systems as the result of the UVIT detection limit.
This results in an intrinsic binary fraction of 18.8\,$\pm$\,1.5\,\%, in the range $0.3 < q < 1.0$.

When comparing intrinsic binary fractions of stellar populations, it is fundamental to consider the range of orbital configurations over which the observations are sensitive.
We place constraints on the range of orbital periods by considering that the physical size of the RSG places a lower limit on the orbital period of around $\log P[{\rm day}]\sim3$.
The upper limit can be estimated using the XMD of 0.4", which at the distance of the SMC corresponds to an orbital period of $\log P[{\rm day}]\sim8$.

Table~\ref{tb:RSG-bf} compares the RSG binary fractions that have been determined for Local Group galaxies.
We find excellent agreement with the LMC study of~\citet{2020ApJ...900..118N} and the lower limit placed by~\citet{2021MNRAS.502.4890D}.
See~\citet{2022MNRAS.513.5847P} for further comparisons at other evolutionary stages.

\begin{table}
  \begin{center}
  \caption{Summary of the determination of the RSG multiplicity fraction in Local Group galaxies.}
  \label{tb:RSG-bf}
 {\scriptsize
  \begin{tabular}{lccl}\hline 
{\bf Galaxy} & {\bf Multiplicity fraction} & {\bf Orbital configuration range} & {\bf Reference} \\ 
& {\bf(\%)} \\ \hline
SMC          & 18.8\,$\pm$\,1.5 & $3 < \log P[{\rm days}] < 8$ & \citet{2022MNRAS.513.5847P} \\
                            &   & $0.3 < q < 1.0$\\
SMC          & $>$15 & -- & \citet{2021MNRAS.502.4890D} \\
NGC\,330$^1$   & 30\,$\pm$\,10 & $2.3 < \log P [\rm days] < 4.3$& {\citet{2020A&A...635A..29P}} \\
                            &   & $0.1 < q < 1.0$ \\ \hline
LMC          & 19.5$^{+7.6}_{-6.7}$ & $0.1 < q < 1.0^2$ & \citet{2020ApJ...900..118N} \\ 
LMC          & $>$15 & -- & \citet{2021MNRAS.502.4890D} \\
30 Dor$^3$   & 30\,$\pm$\,10 & $3.3 < \log P [\rm days] < 4.3$ & {\citet{p19}} \\ 
                           & & $0.3 < q < 1.0$\\ \hline
MW           & 27\,$\pm$\,10 & $\log P < 2.9^{5}$ &  {\citet{burki}} \\ 
                             & & $q > 0.1$ \\\hline
M31          & 33.5$^{+8.6}_{-5.0}$ & $0.1 < q < 1.0^2$ & \citet{2021ApJ...908...87N} \\ \hline
M33          & 15-40$^4$ & $0.1 < q < 1.0^2$ & \citet{2021ApJ...908...87N} \\ \hline
  \end{tabular}
  }
 \end{center}
\vspace{1mm}
 \scriptsize{
 {\it Notes:}\\
  $^1$Young massive cluster in the SMC. \\
  $^2$Neugent (priv. comm.)\\
  $^3$LMC star-forming region. As part of the VLT-FLAMES Tarantula Survey~\citet{2011A&A...527A..50E}.\\
  $^4$\citet{2021ApJ...908...87N} find a metallicity dependence of the RSG binary fraction\\
  $^5$\citet{burki} consider stars down to 5\,M$_\odot$, which results in small allowed orbital periods. 
  }
\end{table}
\section{Conclusions}
In this article we present the key results from~\citet{2022MNRAS.513.5847P}, where we aim to identify and study RSG binary systems in the SMC using a newly available UV photometric survey.
We have demonstrated that using UV photometry is an extremely effective method to not only identify RSG binary systems, but to characterise their companions, which is something that until now, has only been possible for a very small number of Galactic systems.

The source list of RSGs that we have compiled consists of 861 high-probability RSGs.
From a cross-match with UV photometry we identify 88 SMC RSG binary systems and we determine stellar parameters for both components in all systems.

By studying the mass-ratios of SMC RSGs we find a flat distribution best describes the observations over the range $0.3 < q < 1.0$.
We use this result to simulate the main observing bias of the UVIT survey to determine the intrinsic RSG multiplicity fraction to be 18.8\,$\pm$\,1.5\,\%, for mass-ratios in the range $0.3 < q < 1.0$ and orbital periods approximately in the range $3 < \log P[{\rm days}] < 8$.
This result is compared to other measurements of the RSG multiplicity fraction in Local Group galaxies, where we find excellent agreement.


\section*{Acknowledgements}
LRP acknowledges the support of the Generalitat Valenciana through the grant APOSTD/2020/247.
This research is partially supported by the Spanish Government under grant PGC2018-093741-B-C21 (MICIU/AEI/FEDER, UE).
DJL acknowledges support from the Spanish Government Ministerio de Ciencia, Innovaci\'on y Universidades through grants PGC-2018-091 3741-B-C22 and from the Canarian Agency for Research, Innovation and Information Society (ACIISI), of the Canary Islands Government, and the European Regional Development Fund (ERDF), under grant with reference ProID2017010115.

\bibliographystyle{apj}
\bibliography{./refs}

\begin{thebibliography}{31}
\expandafter\ifx\csname natexlab\endcsname\relax\def\natexlab#1{#1}\fi

\bibitem[{{Beasor} {et~al.}(2019){Beasor}, {Davies}, {Smith}, \&
  {Bastian}}]{2019MNRAS.486..266B}
{Beasor}, E.~R., {Davies}, B., {Smith}, N., \& {Bastian}, N. 2019, \mnras, 486,
  266

\bibitem[{{Bianchi} \& {Shiao}(2020)}]{2020ApJS..250...36B}
{Bianchi}, L., \& {Shiao}, B. 2020, \apjs, 250, 36

\bibitem[{{Britavskiy} {et~al.}(2019){Britavskiy}, {Lennon}, {Patrick},
  {Evans}, {Herrero}, {Langer}, {van Loon}, {Clark}, {Schneider}, {Almeida},
  {Sana}, {de Koter}, \& {Taylor}}]{b19}
{Britavskiy}, N., {et~al.} 2019, A\&A, 624, A128

\bibitem[{{Burki} \& {Mayor}(1983)}]{burki}
{Burki}, G., \& {Mayor}, M. 1983, \aap, 124, 256

\bibitem[{{Choi} {et~al.}(2016){Choi}, {Dotter}, {Conroy}, {Cantiello},
  {Paxton}, \& {Johnson}}]{2016ApJ...823..102C}
{Choi}, J., {Dotter}, A., {Conroy}, C., {Cantiello}, M., {Paxton}, B., \&
  {Johnson}, B.~D. 2016, \apj, 823, 102

\bibitem[{{Davies} {et~al.}(2013){Davies}, {Kudritzki}, {Plez}, {Trager},
  {Lancon}, {Gazak}, {Bergemann}, {Evans}, \&
  {Chiavassa}}]{2013ApJ...767....3D}
{Davies}, B., {et~al.} 2013, ApJ, 767, 3

\bibitem[{{Dorda} \& {Patrick}(2021)}]{2021MNRAS.502.4890D}
{Dorda}, R., \& {Patrick}, L.~R. 2021, \mnras, 502, 4890

\bibitem[{{Dotter}(2016)}]{2016ApJS..222....8D}
{Dotter}, A. 2016, \apjs, 222, 8

\bibitem[{{Eldridge} {et~al.}(2008){Eldridge}, {Izzard}, \&
  {Tout}}]{2008MNRAS.384.1109E}
{Eldridge}, J.~J., {Izzard}, R.~G., \& {Tout}, C.~A. 2008, \mnras, 384, 1109

\bibitem[{{Evans} {et~al.}(2011){Evans}, {Davies}, {Kudritzki}, {Puech},
  {Yang}, {Cuby}, {Figer}, {Lehnert}, {Morris}, \&
  {Rousset}}]{2011A&A...527A..50E}
{Evans}, C.~J., {et~al.} 2011, A\&A, 527, A50

\bibitem[{{Gaia Collaboration} {et~al.}(2021){Gaia Collaboration}, {Brown},
  {Vallenari}, {Prusti}, {de Bruijne}, {Babusiaux}, {Biermann}, {Creevey},
  {Evans}, {Eyer}, {Hutton}, {Jansen}, {Jordi}, {Klioner}, {Lammers},
  {Lindegren}, {Luri}, {Mignard}, {Panem}, {Pourbaix}, {Randich}, {Sartoretti},
  {Soubiran}, {Walton}, {Arenou}, {Bailer-Jones}, {Bastian}, {Cropper},
  {Drimmel}, {Katz}, {Lattanzi}, {van Leeuwen}, {Bakker}, {Cacciari},
  {Casta{\~n}eda}, {De Angeli}, {Ducourant}, {Fabricius}, {Fouesneau},
  {Fr{\'e}mat}, {Guerra}, {Guerrier}, {Guiraud}, {Jean-Antoine Piccolo},
  {Masana}, {Messineo}, {Mowlavi}, {Nicolas}, {Nienartowicz}, {Pailler},
  {Panuzzo}, {Riclet}, {Roux}, {Seabroke}, {Sordo}, {Tanga}, {Th{\'e}venin},
  {Gracia-Abril}, {Portell}, {Teyssier}, {Altmann}, {Andrae}, {Bellas-Velidis},
  {Benson}, {Berthier}, {Blomme}, {Brugaletta}, {Burgess}, {Busso}, {Carry},
  {Cellino}, {Cheek}, {Clementini}, {Damerdji}, {Davidson}, {Delchambre},
  {Dell'Oro}, {Fern{\'a}ndez-Hern{\'a}ndez}, {Galluccio}, {Garc{\'\i}a-Lario},
  {Garcia-Reinaldos}, {Gonz{\'a}lez-N{\'u}{\~n}ez}, {Gosset}, {Haigron},
  {Halbwachs}, {Hambly}, {Harrison}, {Hatzidimitriou}, {Heiter},
  {Hern{\'a}ndez}, {Hestroffer}, {Hodgkin}, {Holl}, {Jan{\ss}en}, {Jevardat de
  Fombelle}, {Jordan}, {Krone-Martins}, {Lanzafame}, {L{\"o}ffler}, {Lorca},
  {Manteiga}, {Marchal}, {Marrese}, {Moitinho}, {Mora}, {Muinonen}, {Osborne},
  {Pancino}, {Pauwels}, {Petit}, {Recio-Blanco}, {Richards}, {Riello},
  {Rimoldini}, {Robin}, {Roegiers}, {Rybizki}, {Sarro}, {Siopis}, {Smith},
  {Sozzetti}, {Ulla}, {Utrilla}, {van Leeuwen}, {van Reeven}, {Abbas}, {Abreu
  Aramburu}, {Accart}, {Aerts}, {Aguado}, {Ajaj}, {Altavilla}, {{\'A}lvarez},
  {{\'A}lvarez Cid-Fuentes}, {Alves}, {Anderson}, {Anglada Varela}, {Antoja},
  {Audard}, {Baines}, {Baker}, {Balaguer-N{\'u}{\~n}ez}, {Balbinot}, {Balog},
  {Barache}, {Barbato}, {Barros}, {Barstow}, {Bartolom{\'e}}, {Bassilana},
  {Bauchet}, {Baudesson-Stella}, {Becciani}, {Bellazzini}, {Bernet}, {Bertone},
  {Bianchi}, {Blanco-Cuaresma}, {Boch}, {Bombrun}, {Bossini}, {Bouquillon},
  {Bragaglia}, {Bramante}, {Breedt}, {Bressan}, {Brouillet}, {Bucciarelli},
  {Burlacu}, {Busonero}, {Butkevich}, {Buzzi}, {Caffau}, {Cancelliere},
  {C{\'a}novas}, {Cantat-Gaudin}, {Carballo}, {Carlucci}, {Carnerero},
  {Carrasco}, {Casamiquela}, {Castellani}, {Castro-Ginard}, {Castro Sampol},
  {Chaoul}, {Charlot}, {Chemin}, {Chiavassa}, {Cioni}, {Comoretto}, {Cooper},
  {Cornez}, {Cowell}, {Crifo}, {Crosta}, {Crowley}, {Dafonte}, {Dapergolas},
  {David}, {David}, {de Laverny}, {De Luise}, {De March}, {De Ridder}, {de
  Souza}, {de Teodoro}, {de Torres}, {del Peloso}, {del Pozo}, {Delbo},
  {Delgado}, {Delgado}, {Delisle}, {Di Matteo}, {Diakite}, {Diener},
  {Distefano}, {Dolding}, {Eappachen}, {Edvardsson}, {Enke}, {Esquej}, {Fabre},
  {Fabrizio}, {Faigler}, {Fedorets}, {Fernique}, {Fienga}, {Figueras},
  {Fouron}, {Fragkoudi}, {Fraile}, {Franke}, {Gai}, {Garabato},
  {Garcia-Gutierrez}, {Garc{\'\i}a-Torres}, {Garofalo}, {Gavras}, {Gerlach},
  {Geyer}, {Giacobbe}, {Gilmore}, {Girona}, {Giuffrida}, {Gomel}, {Gomez},
  {Gonzalez-Santamaria}, {Gonz{\'a}lez-Vidal}, {Granvik},
  {Guti{\'e}rrez-S{\'a}nchez}, {Guy}, {Hauser}, {Haywood}, {Helmi}, {Hidalgo},
  {Hilger}, {H{\l}adczuk}, {Hobbs}, {Holland}, {Huckle}, {Jasniewicz},
  {Jonker}, {Juaristi Campillo}, {Julbe}, {Karbevska}, {Kervella}, {Khanna},
  {Kochoska}, {Kontizas}, {Kordopatis}, {Korn}, {Kostrzewa-Rutkowska},
  {Kruszy{\'n}ska}, {Lambert}, {Lanza}, {Lasne}, {Le Campion}, {Le Fustec},
  {Lebreton}, {Lebzelter}, {Leccia}, {Leclerc}, {Lecoeur-Taibi}, {Liao},
  {Licata}, {Lindstr{\o}m}, {Lister}, {Livanou}, {Lobel}, {Madrero Pardo},
  {Managau}, {Mann}, {Marchant}, {Marconi}, {Marcos Santos}, {Marinoni},
  {Marocco}, {Marshall}, {Martin Polo}, {Mart{\'\i}n-Fleitas}, {Masip},
  {Massari}, {Mastrobuono-Battisti}, {Mazeh}, {McMillan}, {Messina},
  {Michalik}, {Millar}, {Mints}, {Molina}, {Molinaro}, {Moln{\'a}r},
  {Montegriffo}, {Mor}, {Morbidelli}, {Morel}, {Morris}, {Mulone}, {Munoz},
  {Muraveva}, {Murphy}, {Musella}, {Noval}, {Ord{\'e}novic}, {Orr{\`u}},
  {Osinde}, {Pagani}, {Pagano}, {Palaversa}, {Palicio}, {Panahi}, {Pawlak},
  {Pe{\~n}alosa Esteller}, {Penttil{\"a}}, {Piersimoni}, {Pineau}, {Plachy},
  {Plum}, {Poggio}, {Poretti}, {Poujoulet}, {Pr{\v{s}}a}, {Pulone}, {Racero},
  {Ragaini}, {Rainer}, {Raiteri}, {Rambaux}, {Ramos}, {Ramos-Lerate}, {Re
  Fiorentin}, {Regibo}, {Reyl{\'e}}, {Ripepi}, {Riva}, {Rixon}, {Robichon},
  {Robin}, {Roelens}, {Rohrbasser}, {Romero-G{\'o}mez}, {Rowell}, {Royer},
  {Rybicki}, {Sadowski}, {Sagrist{\`a} Sell{\'e}s}, {Sahlmann}, {Salgado},
  {Salguero}, {Samaras}, {Sanchez Gimenez}, {Sanna}, {Santove{\~n}a},
  {Sarasso}, {Schultheis}, {Sciacca}, {Segol}, {Segovia}, {S{\'e}gransan},
  {Semeux}, {Shahaf}, {Siddiqui}, {Siebert}, {Siltala}, {Slezak}, {Smart},
  {Solano}, {Solitro}, {Souami}, {Souchay}, {Spagna}, {Spoto}, {Steele},
  {Steidelm{\"u}ller}, {Stephenson}, {S{\"u}veges}, {Szabados}, {Szegedi-Elek},
  {Taris}, {Tauran}, {Taylor}, {Teixeira}, {Thuillot}, {Tonello}, {Torra},
  {Torra}, {Turon}, {Unger}, {Vaillant}, {van Dillen}, {Vanel}, {Vecchiato},
  {Viala}, {Vicente}, {Voutsinas}, {Weiler}, {Wevers}, {Wyrzykowski}, {Yoldas},
  {Yvard}, {Zhao}, {Zorec}, {Zucker}, {Zurbach}, \&
  {Zwitter}}]{2021A&A...649A...1G}
{Gaia Collaboration} {et~al.} 2021, \aap, 649, A1

\bibitem[{{Gieles} {et~al.}(2010){Gieles}, {Sana}, \& {Portegies
  Zwart}}]{2010MNRAS.402.1750G}
{Gieles}, M., {Sana}, H., \& {Portegies Zwart}, S.~F. 2010, \mnras, 402, 1750

\bibitem[{{Gordon} {et~al.}(2003){Gordon}, {Clayton}, {Misselt}, {Landolt}, \&
  {Wolff}}]{2003ApJ...594..279G}
{Gordon}, K.~D., {Clayton}, G.~C., {Misselt}, K.~A., {Landolt}, A.~U., \&
  {Wolff}, M.~J. 2003, \apj, 594, 279

\bibitem[{{Langer}(2012)}]{Langer12}
{Langer}, N. 2012, ARA\&A, 50, 107

\bibitem[{{Langer} {et~al.}(2020){Langer}, {Sch{\"u}rmann}, {Stoll},
  {Marchant}, {Lennon}, {Mahy}, {de Mink}, {Quast}, {Riedel}, {Sana},
  {Schneider}, {Schootemeijer}, {Wang}, {Almeida}, {Bestenlehner},
  {Bodensteiner}, {Castro}, {Clark}, {Crowther}, {Dufton}, {Evans}, {Fossati},
  {Gr{\"a}fener}, {Grassitelli}, {Grin}, {Hastings}, {Herrero}, {de Koter},
  {Menon}, {Patrick}, {Puls}, {Renzo}, {Sander}, {Schneider}, {Sen}, {Shenar},
  {Sim{\'o}n-D{\'\i}as}, {Tauris}, {Tramper}, {Vink}, \&
  {Xu}}]{2020A&A...638A..39L}
{Langer}, N., {et~al.} 2020, \aap, 638, A39

\bibitem[{{Moe} \& {Di Stefano}(2017)}]{2017ApJS..230...15M}
{Moe}, M., \& {Di Stefano}, R. 2017, ApJS, 230, 15

\bibitem[{{Neugent}(2021)}]{2021ApJ...908...87N}
{Neugent}, K.~F. 2021, \apj, 908, 87

\bibitem[{{Neugent} {et~al.}(2020){Neugent}, {Levesque}, {Massey}, {Morrell},
  \& {Drout}}]{2020ApJ...900..118N}
{Neugent}, K.~F., {Levesque}, E.~M., {Massey}, P., {Morrell}, N.~I., \&
  {Drout}, M.~R. 2020, \apj, 900, 118

\bibitem[{{Neugent} {et~al.}(2018){Neugent}, {Massey}, {Morrell}, {Skiff}, \&
  {Georgy}}]{2018AJ....155..207N}
{Neugent}, K.~F., {Massey}, P., {Morrell}, N.~I., {Skiff}, B., \& {Georgy}, C.
  2018, \aj, 155, 207

\bibitem[{{Patrick} {et~al.}(2021){Patrick}, {Bianchi}, {Dorda}, {Langer},
  {Lennon}, {Negueruela}, \& {Thilker}}]{2021hst..prop16776P}
{Patrick}, L., {Bianchi}, L.~C., {Dorda}, R., {Langer}, N., {Lennon}, D.~J.,
  {Negueruela}, I., \& {Thilker}, D. 2021, {Cool stars with hot companions in
  the Small Magellanic Cloud}, HST Proposal. Cycle 29, ID. \#16776

\bibitem[{{Patrick} {et~al.}(2016){Patrick}, {Evans}, {Davies}, {Kudritzki},
  {H{\'e}nault-Brunet}, {Bastian}, {Lapenna}, \&
  {Bergemann}}]{2016MNRAS.458.3968P}
{Patrick}, L.~R., {Evans}, C.~J., {Davies}, B., {Kudritzki}, R.-P.,
  {H{\'e}nault-Brunet}, V., {Bastian}, N., {Lapenna}, E., \& {Bergemann}, M.
  2016, \mnras, 458, 3968

\bibitem[{{Patrick} {et~al.}(2022){Patrick}, {Thilker}, {Lennon}, {Bianchi},
  {Schootemeijer}, {Dorda}, {Langer}, \& {Negueruela}}]{2022MNRAS.513.5847P}
{Patrick}, L.~R., {Thilker}, D., {Lennon}, D.~J., {Bianchi}, L.,
  {Schootemeijer}, A., {Dorda}, R., {Langer}, N., \& {Negueruela}, I. 2022,
  \mnras, 513, 5847

\bibitem[{{Patrick} {et~al.}(2019){Patrick}, {Lennon}, {Britavskiy}, {Evans},
  {Sana}, {Taylor}, {Herrero}, {Almeida}, {Clark}, {Gieles}, {Langer},
  {Schneider}, \& {van Loon}}]{p19}
{Patrick}, L.~R., {et~al.} 2019, \aap, 624, A129

\bibitem[{{Patrick} {et~al.}(2020){Patrick}, {Lennon}, {Evans}, {Sana},
  {Bodensteiner}, {Britavskiy}, {Dorda}, {Herrero}, {Negueruela}, \& {de
  Koter}}]{2020A&A...635A..29P}
---. 2020, \aap, 635, A29

\bibitem[{{Sana} {et~al.}(2012){Sana}, {de Mink}, {de Koter}, {Langer},
  {Evans}, {Gieles}, {Gosset}, {Izzard}, {Le Bouquin}, \&
  {Schneider}}]{2012Sci...337..444S}
{Sana}, H., {et~al.} 2012, Science, 337, 444

\bibitem[{{Schneider} {et~al.}(2019){Schneider}, {Ohlmann}, {Podsiadlowski},
  {R{\"o}pke}, {Balbus}, {Pakmor}, \& {Springel}}]{2019Natur.574..211S}
{Schneider}, F. R.~N., {Ohlmann}, S.~T., {Podsiadlowski}, P., {R{\"o}pke},
  F.~K., {Balbus}, S.~A., {Pakmor}, R., \& {Springel}, V. 2019, \nat, 574, 211

\bibitem[{{Schneider} {et~al.}(2014){Schneider}, {Izzard}, {de Mink}, {Langer},
  {Stolte}, {de Koter}, {Gvaramadze}, {Hu{\ss}mann}, {Liermann}, \&
  {Sana}}]{2014ApJ...780..117S}
{Schneider}, F.~R.~N., {et~al.} 2014, \apj, 780, 117

\bibitem[{{Schootemeijer} {et~al.}(2019){Schootemeijer}, {Langer}, {Grin}, \&
  {Wang}}]{2019A&A...625A.132S}
{Schootemeijer}, A., {Langer}, N., {Grin}, N.~J., \& {Wang}, C. 2019, \aap,
  625, A132

\bibitem[{{Sen} {et~al.}(2022){Sen}, {Langer}, {Marchant}, {Menon}, {de Mink},
  {Schootemeijer}, {Sch{\"u}rmann}, {Mahy}, {Hastings}, {Nathaniel}, {Sana},
  {Wang}, \& {Xu}}]{2022A&A...659A..98S}
{Sen}, K., {et~al.} 2022, \aap, 659, A98

\bibitem[{{Tabernero} {et~al.}(2018){Tabernero}, {Dorda}, {Negueruela}, \&
  {Gonz{\'a }lez-Fern{\'a}ndez}}]{2018MNRAS.476.3106T}
{Tabernero}, H.~M., {Dorda}, R., {Negueruela}, I., \& {Gonz{\'a
  }lez-Fern{\'a}ndez}, C. 2018, \mnras, 476, 3106

\bibitem[{{Yang} {et~al.}(2020){Yang}, {Bonanos}, {Jiang}, {Gao}, {Gavras},
  {Maravelias}, {Wang}, {Chen}, {Tramper}, {Ren}, {Spetsieri}, \&
  {Xue}}]{2020A&A...639A.116Y}
{Yang}, M., {et~al.} 2020, \aap, 639, A116

\end{thebibliography}
\clearpage
\begin{discussion}

\discuss{Bostroem}{
  How do you deal with line of sight coincides and recognise true binaries?
}

\discuss{Patrick}{
  This is an excellent question and something I think is really important to consider. I didn't have time to elaborate on this much in the talk. We assess the probability of chance alignments using a control catalogue. We find that outside of a cross-match distance of 0.5" false positive detections can account for up to 100\% of matches. Within 0.4", the false positive detection rate is effectively negligible. It is only thanks to the high precision of our UV point source catalogue from the UVIT/Astrosat survey that we are able to do this so accurately. 
}

\discuss{Graefener}{
  Are there signatures of wind emission from the RSG in these detections?
  Could these detections in the UV be the result of the RSG wind?
}

\discuss{Patrick}{
  As Goetz points out, in the IUE spectra of Galactic RSG binary systems there are signatures of wind emission from the RSG to varying degrees.
  In the UV photometry that I have shown today, these signatures do not contribute significantly to the flux that we observe and because of that we are confident that the UV flux comes almost exclusively from the hot star companion. 
  We have HST STIS UV spectra from our recent SNAP proposal~\citet{2021hst..prop16776P} that do show evidence of wind emission features that originate from the RSG.
  Single RSGs on the other hand, would be far too faint in the UV to be detected with the UV photometry presented here.
}
\discuss{Stanway}{
  I'm not sure I understood how you get both a temp and luminosity from a single UV point. Can you clarify?
}
\discuss{Patrick}{
  We determine the luminosity from the UVIT magnitudes and bolometric correction from the MIST models. We then assume an age (which is taken from the RSG age) to determine the effective temperatures and masses from the MIST models.
  The grey solid lines in Figure~\ref{fig:HRD} highlight the relationship between effective temperature and luminosity for constant UVIT magnitudes.
  Because these lines are almost perpendicular to the main sequence, by assuming an age, we can do a good job of determining the stellar parameters of the companions despite the observational limitation of only using a single photometric point. In the future we hope to supplement this with further UV imaging. 
}
\end{discussion}

\end{document}